\documentclass{article}

\usepackage{arxiv}

\usepackage{graphicx} 
\usepackage{amssymb}
\usepackage{pifont}
\newcommand{\cmark}{\ding{51}}%
\newcommand{\xmark}{\ding{55}}%
\usepackage{cite}
\usepackage{url}
\usepackage[hidelinks]{hyperref}
\usepackage{pgfplots}
\usepackage{subfig}
\usepackage{multirow}
\usepackage[linesnumbered,ruled,vlined]{algorithm2e}
\usepackage{rotating}

\pgfplotsset{compat=1.18}
\newcommand{\wrap}[1]{\begin{tabular}{@{}c@{}}#1\end{tabular}}

\usetikzlibrary{external}
\tikzset{external/mode=graphics if exists}

\usepackage[title]{appendix}

\title{FLsim: A Modular and Library-Agnostic Simulation Framework for Federated Learning}

\author{
 Arnab Mukherjee \\
  Dept. of Computer Science \& Engineering\\
  Indian Institute of Technology Patna\\
  Patna, IN 801106 \\
  \texttt{arnab\_2213cs01@iitp.ac.in} \\
   \And
 Raju Halder \\
  Dept. of Computer Science \& Engineering\\
  Indian Institute of Technology Patna\\
  Patna, IN 801106 \\
  \texttt{halder@iitp.ac.in} \\
  \And
 Joydeep Chandra \\
  Dept. of Computer Science \& Engineering\\
  Indian Institute of Technology Patna\\
  Patna, IN 801106 \\
  \texttt{joydeep@iitp.ac.in} \\
  }

\begin{document}
\maketitle

\begin{abstract}
     Federated Learning (FL) has undergone significant development since its inception in 2016, advancing from basic algorithms to complex methodologies tailored to address diverse challenges and use cases. However, research and benchmarking of novel FL techniques against a plethora of established state-of-the-art solutions remain challenging. To streamline this process, we introduce FLsim, a comprehensive FL simulation framework designed to meet the diverse requirements of FL workflows in the literature. FLsim is characterized by its modularity, scalability, resource efficiency, and controlled reproducibility of experimental outcomes. Its easy to use interface allows users to specify customized FL requirements through job configuration, which supports: (a) customized data distributions, ranging from non-independent and identically distributed (non-iid) data to independent and identically distributed (iid) data, (b) selection of local learning algorithms according to user preferences, with complete agnosticism to ML libraries, (c) choice of network topology illustrating communication patterns among nodes, (d) definition of model aggregation and consensus algorithms, and (e) pluggable blockchain support for enhanced robustness. Through a series of experimental evaluations, we demonstrate the effectiveness and versatility of FLsim in simulating a diverse range of state-of-the-art FL experiments. We envisage that FLsim would mark a significant advancement in FL simulation frameworks, offering unprecedented flexibility and functionality for researchers and practitioners alike.
\end{abstract}


\section{Introduction}\label{sec:intro}

Since its introduction by Google in 2016 \cite{fedavg_paper}, Federated Learning (FL) has seen significant development over the years. To address several dominant challenges pertaining to FL, the concept has evolved from simple algorithms like Federated Averaging (FedAvg) \cite{fedavg_paper} to more sophisticated techniques including server-momentum \cite{fedavgm}, client-side regularization \cite{fedprox,moon,scaffold}, server-side optimization \cite{server_opt_fed}, FL with Differential Privacy \cite{client_diffpriv,pcfed-tii}, Personalized FL \cite{fedma,fedper}, Generalized and Clustered FL \cite{grace,tii-cluster}, blockchain-based FL for trusted and robust aggregation \cite{fedrlchain,bcsecagg-tii}, and communication efficient FL \cite{activesemi-tsmcs,fedstream-tsmcs}. Each of these techniques distinguishes itself based on a variety of factors, including the local learning algorithm employed, the aggregation scheme used to generate a global model, the communication topology, and any verification and consensus algorithm. With an extensive array of existing FL solutions available, researching a novel FL technique and benchmarking it against state-of-the-art solutions is a challenge. To streamline this process, the need for a modular, scalable, and resource-efficient federated learning simulation and testing framework is essential.

This paper presents \textsf{FLsim}, an FL simulation framework to achieve the following goals: (1) support diverse FL framework requirements since there exists a plethora of proposals requiring customization at various levels and stages of the FL workflow, (2) enabling the learning over different data distribution across clients, spanning from non-independent and identically distributed (non-iid) (e.g., healthcare), to independent and identically distributed (iid) (e.g., simulated experiments),
(3) achieving complete ML library agnosticism to meet the demands of the diverse community of users preferring one ML library over the others, (4) support diverse network topologies ranging from client-server to decentralized topology, (5) controlled reproducibility of experimental outcomes to easily gauge the effect on experimental outcomes through tweaking and tuning hyperparameters and architectures. (6) scalable enough to support a large number of nodes since real-world FL use cases might scale from siloed data sites to thousands of edge clients. (7) pluggable support for blockchain-based verifiability and consensus for decision-making, for traceable and trusted execution among multiple nodes spread across a network. 

With the modular workflow for defining and implementing FL algorithms provided with \textsf{FLsim}, users need to define the following six critical requirements for its simulation: (1) the dataset and distribution to be followed, (2) the deep learning model and training loop to be used, (3) the aggregation algorithm to be used, (4) the testing workflow to be implemented and the metrics to use, (5) the information that needs to be communicated among the nodes (such as local/global models, additional states such as control variates, etc.), (6) in case of a multi-worker aggregation scenario, what consensus algorithm to use for selection of the global model. The framework automatically handles the rest of the infrastructure and workflow based on these six requirements provided via the \textsf{FLsim} job configuration. 

Unlike existing simulation-oriented libraries like Tensorflow Federated (TFF) \cite{tff} and PySyft \cite{pysyft}, which are only limited to client-server topology and support basic algorithms like FedAvg \cite{fedavg_paper} or FedProx \cite{fedprox}, with execution limited within a single machine, our framework is able to follow any given topology from basic client-server to complex peer-to-peer topologies. \textsf{FLsim} provides a modular and customizable experience out-of-the-box when it comes to defining aggregation and training algorithms, which libraries like FATE \cite{fate} and FedBioMed \cite{fedbiomed} lack, limiting them from algorithmic innovations for open FL problems. Additionally, \textsf{FLsim} provides a tight-knit environment for controlled and reproducible experiments with the support for metrics generation and extensive logging, which existing proposals like Flower \cite{flower}, FedLab \cite{fedlab} and FedML \cite{fedml} lack. Finally, compared to \cite{fedstellar}, \textsf{FLsim} is not limited to only certain decentralized FL workflows and is much more flexible when it comes to supporting multi-worker consensus and support for blockchain-aided decision-making. To this end, there is no exact platform that caters to the diverse requirements when it comes to exploratory research on federated learning, and hence, we embark on developing an FL simulation framework that caters to the vast requirements that encompass to developing FL algorithms and techniques.

Fundamentally, we aim to address the gaps within the existing simulation framework, answering the following research questions: \textbf{RQ1:} Is \textsf{FLsim} able to support the diverse landscape of FL computing paradigms? \textbf{RQ2:} Is \textsf{FLsim} generic enough to be completely ML library agnostic? \textbf{RQ3:} Does the framework support parameter verification and consensus among nodes for the selection of global models in multi-aggregator environments? \textbf{RQ4:} In terms of trust and security, does the framework support platform-agnostic blockchain-based traceability and verifiability of the learning process? \textbf{RQ5:} Is the framework able to support the diverse possible network topologies for FL? \textbf{RQ6:} Being a simulation framework, does the platform support deterministic and reproducible experimental results? \textbf{RQ7:} Is the framework scalable enough and resource-efficient for large-scale deployment in resource-constraint environments and devices?
   
To summarise, this paper presents the following contributions:

\begin{itemize}
    \item We propose \textsf{FLsim}, a modular and robust framework designed for federated learning (FL) simulation. \textsf{FLsim} offers a customizable experience by enabling users to define aggregation and training algorithms while remaining completely agnostic to machine learning libraries. It supports a wide range of network topologies, from basic client-server configurations to intricate peer-to-peer setups. Additionally, \textsf{FLsim} is flexible in accommodating multi-worker consensus and facilitates blockchain-aided decision-making. Furthermore, the framework ensures controlled and reproducible experiments with support for custom metrics and extensive logging.

    \item We provide a detailed overview of the layered architecture employed by \textsf{FLsim}, elucidating the core components of the framework. Additionally, we delineate the workflow for executing FL experiments, elaborating on the intricacies of job configurations and FL strategy definitions.

    \item We conduct comprehensive evaluations of \textsf{FLsim} under diverse experimental scenarios to assess its scalability, efficiency, and modularity. Our experiments encompass various aspects, such as support for different FL proposals, ML library agnosticism, and reproducibility of FL experiments. Through detailed experimental results, we demonstrate the effectiveness and versatility of \textsf{FLsim} in addressing key challenges in FL research and experimentation.
\end{itemize}

The remainder of the paper is structured as follows: A detailed introduction to \textsf{FLsim}, delineating the core components and workflow, is presented in Section \ref{sec:proposal}. Section \ref{sec:implementation} provides a comprehensive overview of the framework's implementation, providing detailed insights into its layered architecture. A thorough experimentation evaluation of the framework under various circumstances is presented in Section \ref{sec:exp-results}, subsequently answering the research questions. Section \ref{sec:rel-work-discussion} presents a comparative analysis of \textsf{FLsim} against some of the state-of-the-art FL simulation and testing platforms, highlighting the distinguishing features of \textsf{FLsim}. Finally, we detail our future scope and conclude our work with Section \ref{sec:conclusion}.

\begin{figure}[t]
    \centering
    \includegraphics[width=1.0\textwidth]{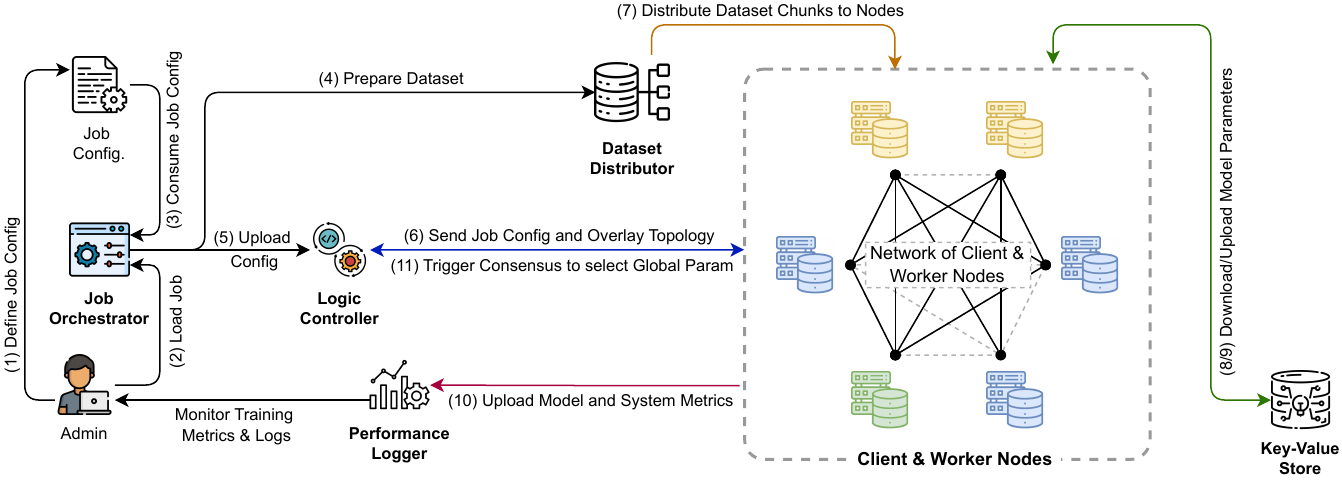}
    \caption{System Workflow of \textsf{FLsim}} 
    \label{fig:distlearn-workflow}
\end{figure}

\section{FLsim: An FL Simulation Framework}\label{sec:proposal}
In this section, we manifest the details of our framework and demonstrate how FL experiments can be scaffolded from the configuration files. The complete system works in synchrony with the six core components of the framework. As depicted in Figure \ref{fig:distlearn-workflow}, the workflow begins with the user defining the job configuration files and loading the job through the Job Orchestrator. Next, the job orchestrator forwards the information to the Dataset Distributor and Logic Controller for scaffolding of datasets and nodes into clients and workers. Once that's done, the nodes download their respective dataset shards and begin the process of federated learning. From time to time, the metrics of model performance and resource usage are reported back to the performance logger for the user to monitor the progress of their experiments. 

Now, let us detail the functions of these six components and the three essential files that the users need to define to scaffold and execute an FL experiment successfully on our platform.

\subsection{FLsim Components}
The core framework consists of six components essential to the functioning of the system:
\begin{enumerate}
    \item \textbf{Job Orchestrator:} The first and foremost component is the Job Orchestrator, which loads the job configuration and FL strategy to scaffold the whole FL workflow. Based on the specification of job configuration (node roles, topology, FL strategy, dataset distribution parameters), it prepares an overlay network specification, along with a bundled package containing the required Python modules, which are forwarded to the Logic Controller for distribution to the respective nodes, and the Dataset Distributor for preparation of the simulated dataset chunks.

    \item \textbf{Logic Controller:} We introduce the Logic Controller to efficiently coordinate and synchronize the client and worker nodes involved in an FL experiment. Acting as the orchestrator of the FL process, the Logic Controller manages the flow of logic and signalling among these nodes, determining when to fetch global parameters, commence training, initiate aggregation, and execute consensus for selecting a global model.

    \item \textbf{Dataset Distributor:} The Dataset Distributor plays a crucial role in the distribution and management of datasets within the system. This component is tasked with archiving and indexing dataset chunks, which are subsequently downloaded by nodes for training or testing purposes. By adhering to the dataset specifications and employing designated distribution algorithms, the distributor efficiently divides datasets into manageable chunks tailored for client and worker nodes.

    \item \textbf{Client / Worker Nodes:} The concept of client and worker nodes varies across different proposals, where nodes may serve exclusively as clients, workers, or perform both roles. Our framework empowers users to designate nodes as clients or workers according to the network definitions outlined in the job configuration. The Logic Controller allocates these roles to specific nodes once the FL job is initialized. 
    
    \item \textbf{Key-Value Store:} The key-value store serves a pivotal role in sharing and storage of states for client and worker nodes, facilitating efficient communication of model parameters and other essential information among them. The key-value store essentially functions as a broker within a pub-sub network. The nodes acting as publishers share their states (including model parameters and additional data) with the key-value store, which eventually shares the same with the subscriber nodes. This setup effectively alleviates communication overhead by avoiding direct node-to-node communication in decentralized FL topologies.

    \item \textbf{Performance Logger and FL-Dashboard:} Since it is essential for extensively visualizing and monitoring the overall learning activities by various nodes and the efficacy of FL models, we implement a performance logging and visualization module baked into the framework. This empowers users to deeply analyze the performance and statistical insights of their FL simulation experiments, discerning how even subtle adjustments impact both the learning trajectory and resource utilization with precision.
   
\end{enumerate}

\subsection{Job Configuration and Strategy Definition}

To scaffold an FL experiment on \textsf{FLsim}, users must define three essential files: the job configuration file, the dataset definition file, and the FL strategy file. Here, we outline the structure and content of these files to clarify their implementation.

\begin{figure}[ht]
	\centering
	\includegraphics[width=1.0\textwidth]{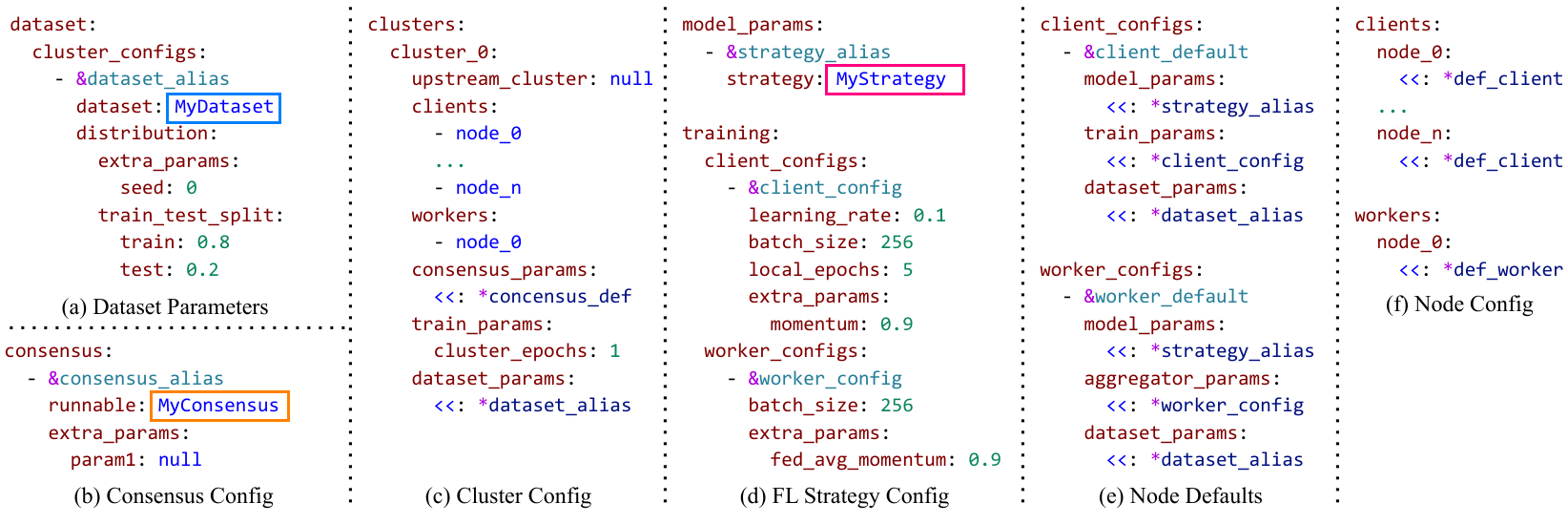}
	\caption{Job Config Structure} 
	\label{fig:job-config-snippet}
\end{figure}

\begin{figure}[t]
	\centering
	\includegraphics[width=0.5\columnwidth]{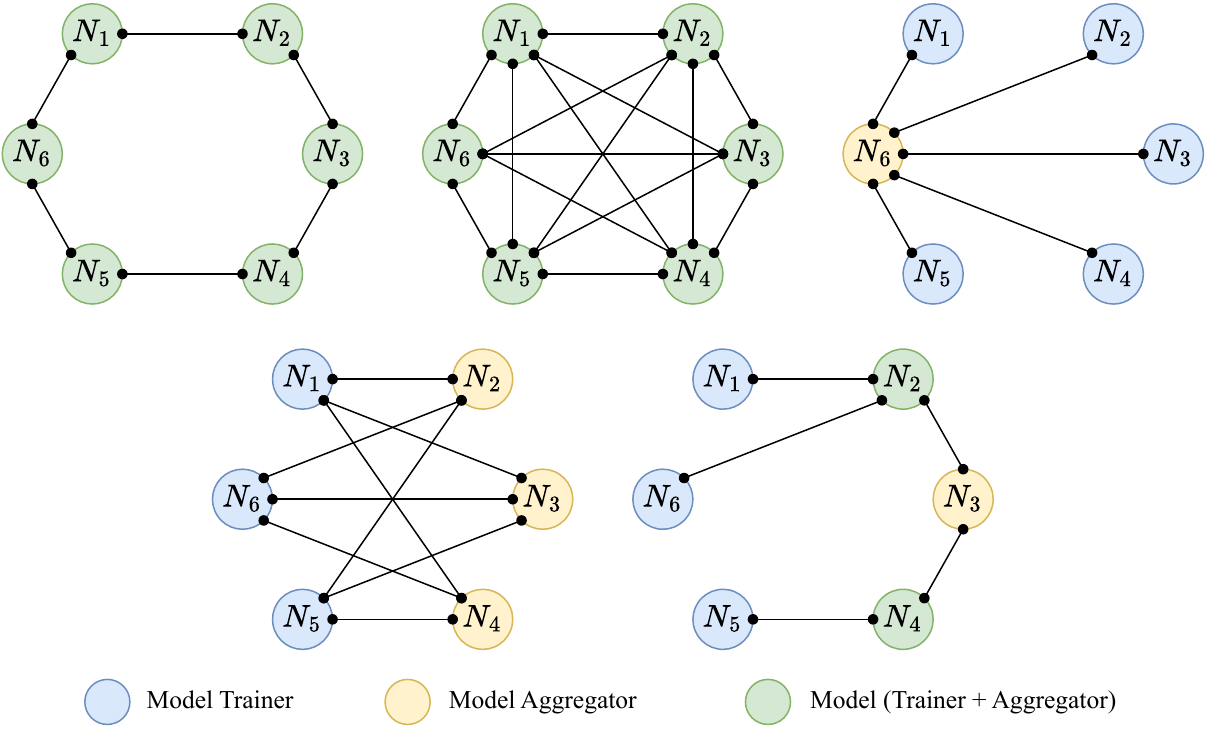}
	\caption{Federated Learning Topologies} 
	\label{fig:fl-topology}
\end{figure}

The job configuration, represented by a \texttt{.yaml} file, serves as the cornerstone for defining the customizable parameters of an FL experiment. The structure of the job configuration is illustrated in Figure \ref{fig:job-config-snippet}. Comprising six integral sections, it includes: (a) dataset parameters, (b) consensus configuration, (c) node overlay topology/cluster configuration, as depicted in Figure \ref{fig:fl-topology}, (d) FL strategy configuration encompassing strategy specifics along with training and aggregation hyperparameters, (e) node default/global configurations, and finally (f) specific-node configurations for each node.

\begin{figure}[ht]
	\centering
	\includegraphics[width=1.0\textwidth]{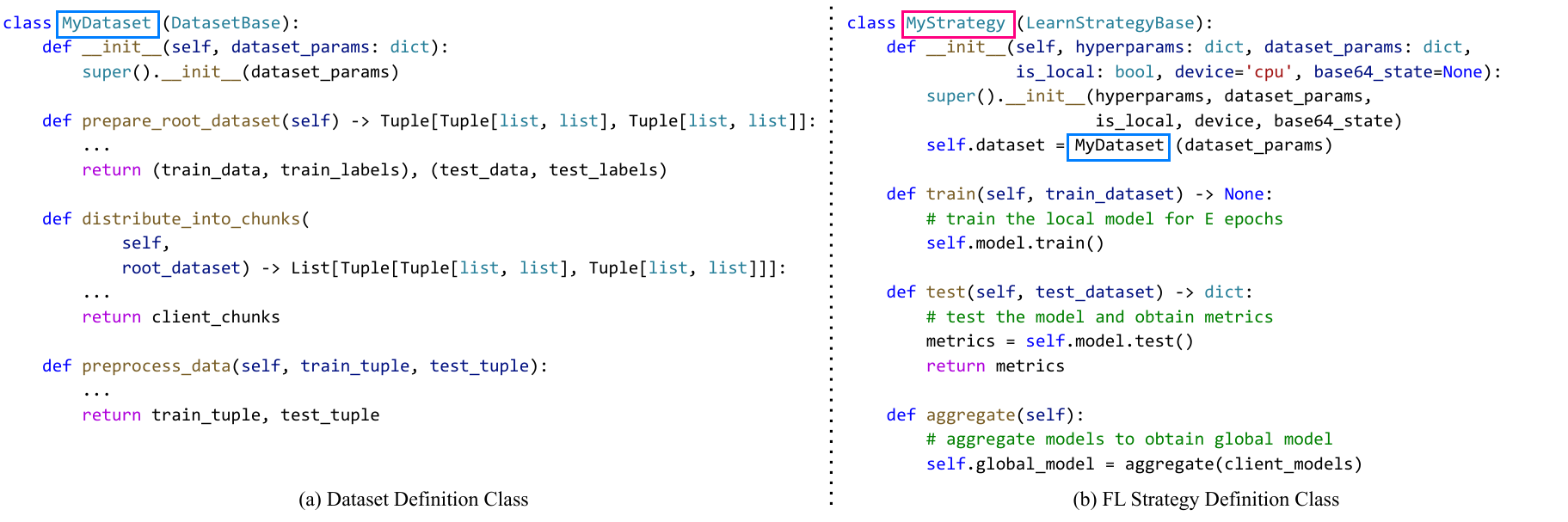}
	\caption{FL Strategy Definition Structure} 
	\label{fig:strategy-snippet}
\end{figure}

In order to empower users with modular and highly customizable configuration options, \textsf{FLsim} facilitates the creation of tailored dataset distributions and complete FL algorithmic strategy according to needs, as depicted in Figure \ref{fig:strategy-snippet}. These definitions are invoked in the job configuration, as denoted by the coloured boxes.

\paragraph{Dataset and Distribution Definition}
As outlined above, to scaffold an FL experiment, users must initially define the dataset and the distribution algorithm to be utilized. To facilitate a modular and customizable experience in dataset handling and distribution, \textsf{FLsim} offers an object-oriented approach. Users can define their dataset class, inherited from the base class of \textsf{FLsim}-Dataset, which incorporates three fundamental methods: (1) the root dataset preparation method \texttt{prepare\_root\_dataset()}, (2) a method for distributing the dataset into client chunks \texttt{distribute\_into\_chunks()}, and (3) a client-side dataset preprocessing method \texttt{preprocess\_data()}. A code snippet illustrating these class methods is presented in Figure \ref{fig:strategy-snippet}a. 
It is worthwhile to highlight that \textsf{FLsim} seamlessly manages the dependencies and logic handling behind the scenes to ensure a smooth experience. Moreover, to streamline control over dataset parameters, the dataset configuration section within the job configuration file houses hyperparameters and file configurations, as illustrated in Figure \ref{fig:job-config-snippet}a.

\paragraph{FLsim FL-Strategy} 
\textsf{FLsim} provides an object-oriented approach to how users can implement their training, testing and aggregation methods. These three methods are part of the \textsf{FLsim} Strategy, which inherits the FL strategy base class. The base class includes the required interfaces along with additional auxiliary methods. The users need to primarily define three methods for their FL Strategy, i.e., (1) the model training method \texttt{train()}, which accepts the training dataset as its arguments, (2) the aggregation method \texttt{aggregate()}, which aggregates the collected client models and creates a new global model, and (3) the testing method \texttt{test()} for testing the local and global model's performance after they are trained or aggregated. The code snipped depicted in Figure \ref{fig:strategy-snippet}b describes the structure of the \textsf{FLsim} Strategy class, along with the three methods the users need to implement. Additionally, along with these three methods, the users also need to initialize the dataset state variable as an instance of the Dataset class they defined prior, as depicted in Figure \ref{fig:strategy-snippet}a.

\subsection{Node Synchronization}
Synchronization among nodes in FL is crucial to ensure that all participating nodes maintain consistent and up-to-date model parameters, enabling effective collaboration and convergence towards a global model despite their distributed nature. To this end, \textsf{FLsim} implements its own synchronization algorithm, which works in unison with the Logic Controller and the nodes (clients and workers). Algorithm \ref{algo:node-sync} presents the logic behind the synchronization among the nodes and the Logic Controller, enabling an efficient and fault-tolerant FL process. The core synchronization is based on two types of signals, i.e., the \texttt{ProcessPhase} and \texttt{NodeStage}. The \texttt{ProcessPhase} denotes at what phase the FL experiment is currently, while the \texttt{NodeStage} denotes at what stage the \textsf{FLsim} nodes are currently. The \texttt{ProcessPhase} has three discrete values, i.e., \texttt{ProcessPhase} $\in \{0,1,2\}$, where 0 = ``System Initializing'', 1 = ``In Local Learning'', 2 = ``In Model Aggergation''. Similarly, the \texttt{NodeStage} $\in \{0,1,2,3,4\}$, where 0 = ``Nodes not Ready'', 1 = ``Nodes Ready for Job'', 2 = ``Nodes Ready with Dataset'', 3 = ``Clients busy in Training'' for clients and ``Workers busy in Aggregation'' for workers, and 4 = ``Clients Waiting for Next Round'' for clients and ``Aggregation Complete'' for workers.

\begin{algorithm}[!ht]
\footnotesize
\SetAlgoLined
\textbf{set} \texttt{ProcessPhase} $\leftarrow$ 0\;
\textbf{set} \texttt{NodeStage} $\leftarrow$ $(0, 0)$\;
\textbf{set} \texttt{DownloadJobConfig} $\leftarrow$ True\;
\For{\textnormal{node} $\in$ \textnormal{(clients + workers)}}{
    node.\texttt{updateNodeStatus(1)}\;
}
\textbf{wait-until} \texttt{all\_nodes\_in\_stage}(1)\;
\textbf{set} \texttt{DownloadDataset} $\leftarrow$ True\;
\For{\textnormal{node} $\in$ \textnormal{(clients + workers)}}{
    node.\texttt{downloadDatasetChunk()}\;
    node.\texttt{updateNodeStatus(2)}\;
}
\textbf{wait-until} \texttt{all\_nodes\_in\_stage}(2)\;
\textbf{set} \texttt{GlobalRound} $\leftarrow$ 1\;
\textbf{set} \texttt{GlobalParam} $\leftarrow$ \texttt{initRandomModel()}\;
\While{true}{
    \For{\textnormal{client} $\in$ \textnormal{clients}}{
        client.\texttt{waitForProcessPhase(1)}\;
    }
    \For{\textnormal{worker} $\in$ \textnormal{workers}}{
        worker.\texttt{waitForProcessPhase(2)}\;
    }
    \textbf{set} \texttt{ProcessPhase} $\leftarrow$ 1\;
    \For{\textnormal{client} $\in$ \textnormal{clients}}{
        \textbf{set} \texttt{LocalParam} $\leftarrow$ \texttt{downloadGlobalParam()}\;
        client.\texttt{updateNodeStatus(3)}\;
    }   
    \textbf{wait-until} \texttt{all\_clients\_in\_stage}(3) $\lor$ \texttt{timeout}()\;
    \textbf{emit} ``Clients are busy in local training.''\;
    \For{\textnormal{client} $\in$ \textnormal{clients}}{
        client.\texttt{trainLocally()}\;
        client.\texttt{uploadTrainedModel()}\;
        client.\texttt{updateNodeStatus(4)}\;
        client.\texttt{waitForProcessPhase(2)}
    }
    \textbf{wait-until} \texttt{all\_clients\_in\_stage}(4) $\lor$ \texttt{timeout}()\;
    \textbf{emit} ``Clients are waiting for next round.''\;
    \textbf{set} \texttt{ProcessPhase} $\leftarrow$ 2\;
    \For{\textnormal{worker} $\in$ \textnormal{workers}}{
        \textbf{set} \texttt{ClientParams} $\leftarrow$ \texttt{downloadClientParams()}\;
        worker.\texttt{updateNodeStatus(3)}\;
    }
    \textbf{wait-until} \texttt{all\_workers\_in\_stage}(3) $\lor$ \texttt{timeout}()\;
    \textbf{emit} ``Workers busy in model aggregation.''\;
    \For{\textnormal{worker} $\in$ \textnormal{workers}}{
        worker.\texttt{aggregateParams(ClientParams)}\;
        worker.\texttt{uploadAggregatedModel()}\;
        worker.\texttt{updateNodeStatus(4)}\;
    }
    \textbf{wait-until} \texttt{all\_workers\_in\_stage}(4) $\lor$ (\texttt{timeout}() $\land$ \texttt{AggregatedParams} $\ge$ 1)\;
    \textbf{emit} ``Received aggregated params''\;
    \textbf{set} \texttt{GlobalParam} $\leftarrow$ \texttt{execConcensus()}\;
    \textbf{set} \texttt{GlobalRound} $\leftarrow$ \texttt{GlobalRound}$+ 1$\;
    \If{\textnormal{\texttt{GlobalRound}} $>$ \textnormal{\texttt{TotalRounds}}}{\textbf{break}}
}
\caption{\textsf{FLsim} Node Synchronization}
\label{algo:node-sync}
\end{algorithm}

\subsection{Pluggable Blockchain Integration}
Since there is a need to support and simulate blockchain-based federated learning (BCFL) solutions, which require a blockchain to delegate some part of the decision-making and computation to a blockchain, \textsf{FLsim} provides a pluggable blockchain API for users to seamlessly connect any blockchain platform, satisfying \textbf{RQ4}. However, due to the existence of a plethora of blockchain platforms with their own set of pros and cons, to this end, \textsf{FLsim} provides out-of-the-box support for Ethereum and Hyperledger Fabric as blockchain platforms to simulate BCFL experiments. Observe that, in order to avail the support of any new blockchain platform, users need to define the following three key functionalities: (1) a wrapper on the \textsf{FLsim} Blockchain API (for the new blockchain platform), (2) the smart contracts for that specific blockchain, and (3) the auto-orchestration script for automatically orchestrating the blockchain nodes (which is an optional step, and only required when using a private blockchain network). By deploying appropriate smart contracts, one can achieve the key benefits of blockchain support, including: (1) model parameter verification, (2) traceability and audatibility of decision-making, (3) global model provenance, (4) node reputation score maintenance, and (5) attack prevention.

\subsection{Aggregation Consensus}
\begin{figure}[ht]
    \centering
    \includegraphics[width=0.6\columnwidth]{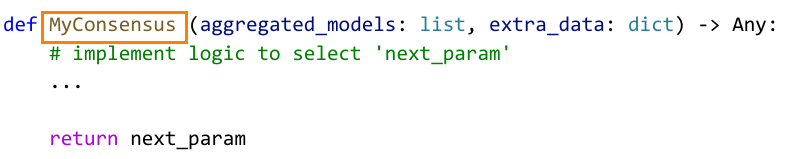}
    \caption{Consensus Method Outline} 
    \label{fig:consensus-snippet}
\end{figure}

In order to support multi-worker FL environments \cite{fedrlchain,deepchain}, \textsf{FLsim} enables users to implement and use consensus mechanisms to decide upon the next global parameter. This consensus mechanism is supported directly through the Logic Controller or can be delegated to a blockchain accompanied by \textsf{FLsim}. In case the consensus process is delegated to a blockchain, the smart contract hosting the consensus logic is executed, and an appropriate global model is selected for the next round. Figure \ref{fig:consensus-snippet} depicts the outline of the function definition to implement a consensus algorithm. The function, whether implemented through the Logic Controller interface or as blockchain smart contracts, accepts the two arguments: (1) a collection of aggregated model parameters and (2) a dictionary containing additional hyperparameters. Based on the logic within the function body, a single model parameter will be returned to be considered as the next global model. Further, to load and execute the consensus algorithm, the consensus name (outlined in orange) is also manifested in the \texttt{.yaml} job configuration, as depicted n Figure \ref{fig:job-config-snippet}b. 

\begin{figure}[ht]
    \centering
    \includegraphics[width=0.7\columnwidth]{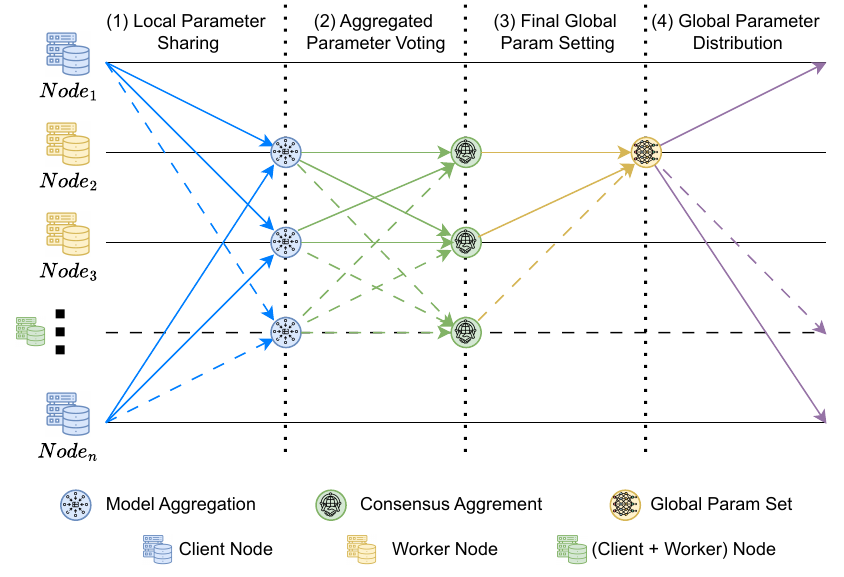}
    \caption{Consensus Workflow} 
    \label{fig:consensus-workflow}
\end{figure}

The overall consensus pipeline for \textsf{FLsim} follows a 4-phase process, where clients and worker nodes participate in one or multiple phases to decide upon a final global model parameter for the next round of local training. As represented in Figure \ref{fig:consensus-workflow}, this process consists of the following four phases: (1) \textbf{Local Parameter Sharing}: Here, the client nodes, after training, share their local model parameters to all of the participating worker nodes for aggregation. (2) \textbf{Aggregated Parameter Voting}: Once the workers have aggregated the received local model parameters, they share the aggregated model parameters (i.e., their hash) among themselves to initiate the consensus process. (3) \textbf{Final Global Parameter Setting}: At this stage, the consensus among the worker nodes is achieved, and a final global model parameter is selected for the next round of federated training. (4) \textbf{Global Parameter Distribution}: Now that a global model is selected, the new model parameter is distributed to all nodes acting as client nodes.

\section{FLsim Implementation}\label{sec:implementation}
In this section, we manifest the implementation details of \textsf{FLsim}. The framework is implemented in Python, with blockchain compatibility provided by individual technology stacks of the respective blockchains. The overall communication among the components of the framework is achieved through REST API calls implemented using the Flask microweb framework. The source code of \textsf{FLsim} is available on GitHub\footnote{\url{https://github.com/mukherjeearnab/FLsim}}.

\begin{figure}[ht]
	\centering
	\includegraphics[width=1.0\textwidth]{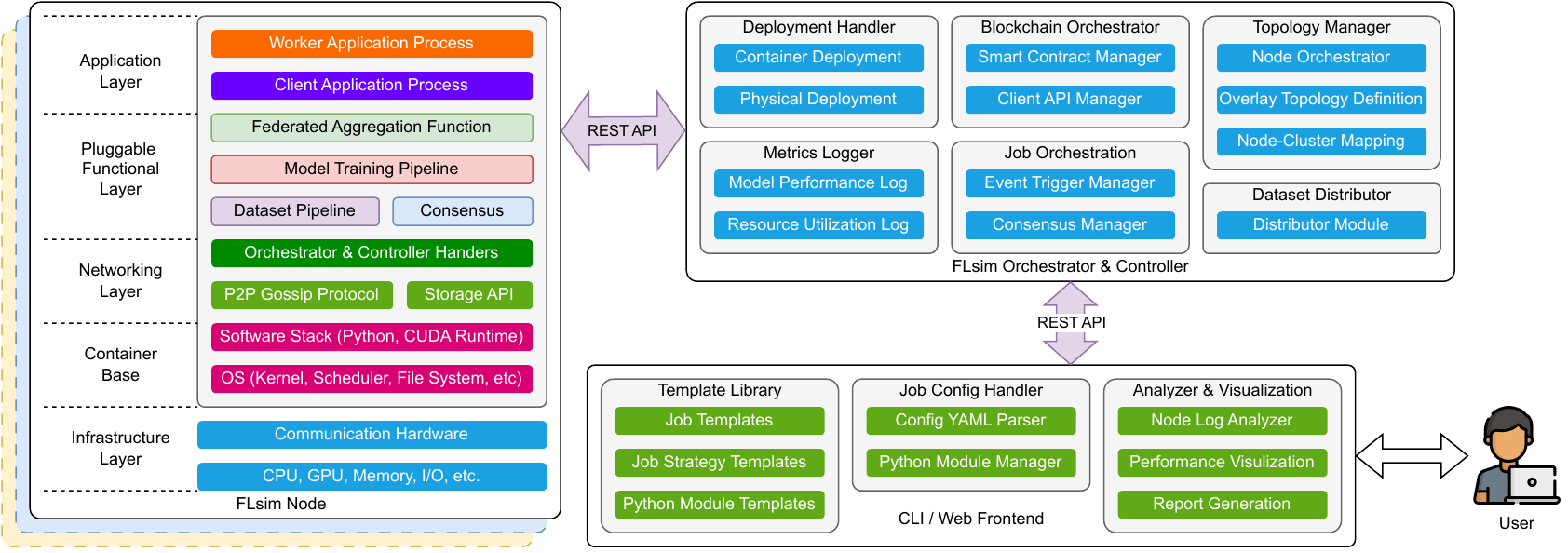}
	\caption{\textsf{FLsim} Framework Overall Architecture} 
	\label{fig:component-diagram}
\end{figure}

A detailed presentation of the overall architecture of the \textsf{FLsim} Framework is presented in Figure \ref{fig:component-diagram}. On the left of the figure, the layers of the \textsf{FLsim} node are represented. These layers work in unison to support the client and worker functionalities during FL simulations. These nodes can be deployed in required numbers to constitute an FL network of clients and workers. Next, there is the \textsf{FLsim} Orchestrator and Controller, which helps the nodes to operate in synchrony and orchestrate the complete FL process. Finally, the CLI/Web frontend is implemented so that the end-user can scaffold and execute federated learning experiments easily.

The \textsf{FLsim} node is comprised of five layers, starting with the infrastructure layer, which is barebones hardware and communication infrastructure. On top of that, the container base is deployed, which serves as the runtime for the OS, Python and CUDA runtimes. Next, we have the networking layer, which hosts the communication modules for the P2P gossip protocol, storage API, and the orchestration and controller handlers. On top of that, the pluggable and modular functional layer is implemented, which includes the dataset pipeline for loading and management of datasets, the consensus protocol for multi-worker strategies, the model training pipeline, and the federated aggregation module. The final and top-most layer is the application layer, which hosts the client and worker application workflow logic.

\section{Experimental Evaluation}\label{sec:exp-results}

In this section, we conduct experimental evaluations on our proposed framework under various circumstances to test the practicality and agility in terms of support for (1) flexibility of implementing existing FL frameworks,
(2) different ML libraries, (3) execution of multi-worker/aggregator-based consensus for FL, (4) orchestration of the platform under different network topologies, (5) reproducibility of the FL experiments on different hardware architectures, and (6) large-scale experiments to demonstrate the scalability of our proposed system.

Let us now start with each of these evaluation circumstances answering the respective research questions formulated in Section \ref{sec:intro}:

\subsection{\textbf{RQ1}: Evaluation on Diverse FL Frameworks}
Here we evaluate the agility of our platform to support diversified state-of-the-art FL frameworks, including FedAvg algorithm \cite{fedavg_paper}, Federated Averaging with Momentum (FedAvgM) \cite{fedavgm}, client and server control variate-based SCAFFOLD \cite{scaffold}, model contrastive learning based MOON \cite{moon}, client differential privacy-based FL technique \cite{client_diffpriv}, hierarchical clustering of client parameter-based framework \cite{hierarchical-cluster}, and decentralized FL-based Fedstellar \cite{fedstellar}. These seven frameworks are selected to demonstrate the ability of \textsf{FLsim} to cater towards the diverse needs of FL algorithms. All the frameworks are implemented on the PyTorch library and tested on the CIFAR-10 dataset distributed using the Dirichlet Distribution algorithm, with $\alpha=0.5$. The deep learning model is comprised of three CNN layers and a fully-connected classification head. A total of 10 clients were involved, where batch size and learning rate were set to 64 and 0.001, respectively.

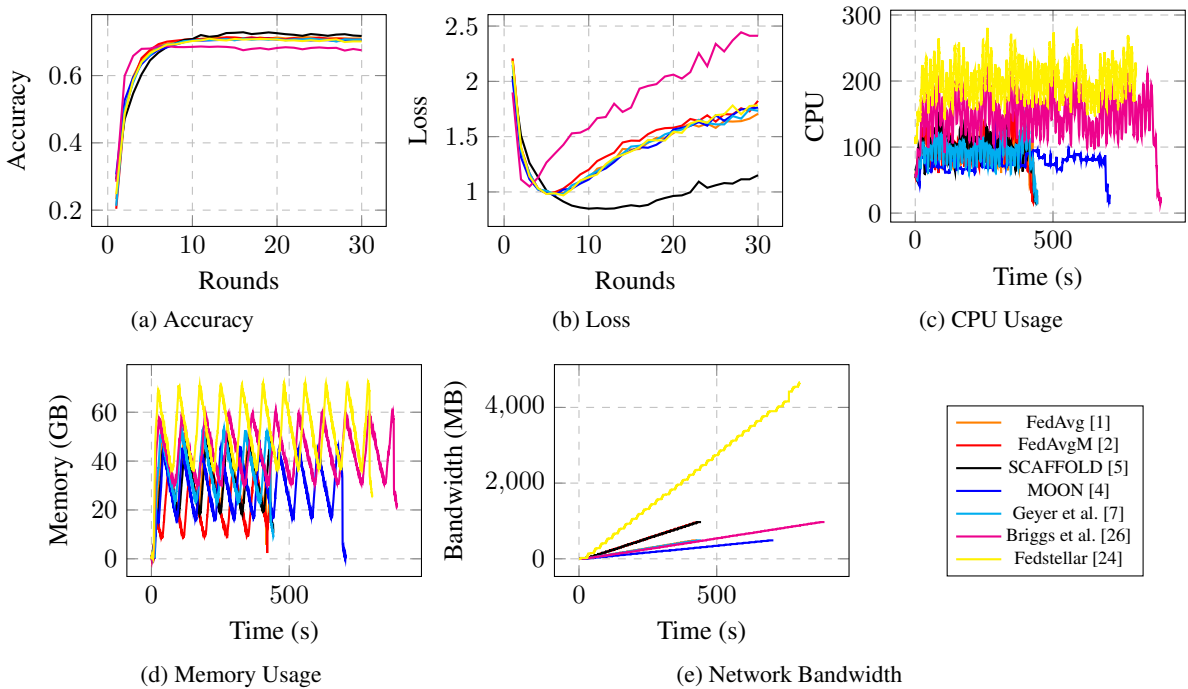
\begin{figure}[ht]%
	\centering
	\subfloat[\centering Accuracy]{{
			\begin{tikzpicture}
				\begin{axis}[xlabel={Rounds},ylabel={Accuracy}, grid=major,grid style={dashed},
					legend pos= south east,legend style={nodes={scale=0.7, transform shape}},
					width=0.333\textwidth,height=4.4cm]
					\addplot[thick, color=orange] table [x=ROUND, y=fedavg, col sep=comma] {data/RQ1/acc.csv};
					\addplot[thick, color=red] table [x=ROUND, y=fedavgm, col sep=comma] {data/RQ1/acc.csv};
					\addplot[thick, color=black] table [x=ROUND, y=scaffold, col sep=comma] {data/RQ1/acc.csv};
					\addplot[thick, color=blue] table [x=ROUND, y=moon, col sep=comma] {data/RQ1/acc.csv};
					\addplot[thick, color=cyan] table [x=ROUND, y=diff, col sep=comma] {data/RQ1/acc.csv};
					\addplot[thick, color=magenta] table [x=ROUND, y=hier, col sep=comma] {data/RQ1/acc.csv};
					\addplot[thick, color=yellow] table [x=ROUND, y=decentral, col sep=comma] {data/RQ1/acc.csv};
				\end{axis}
			\end{tikzpicture}
	}}%
	\subfloat[\centering Loss]{{
			\begin{tikzpicture}
				\begin{axis}[xlabel={Rounds},ylabel={Loss}, grid=major,grid style={dashed},
					legend pos= north east,legend style={nodes={scale=0.7, transform shape}},
					width=0.333\textwidth,height=4.4cm]
					\addplot[thick, color=orange] table [x=ROUND, y=FEDAVG, col sep=comma] {data/RQ1/loss.csv};
					\addplot[thick, color=red] table [x=ROUND, y=FEDAVGM, col sep=comma] {data/RQ1/loss.csv};
					\addplot[thick, color=black] table [x=ROUND, y=SCAFFOLD, col sep=comma] {data/RQ1/loss.csv};
					\addplot[thick, color=blue] table [x=ROUND, y=MOON, col sep=comma] {data/RQ1/loss.csv};
					\addplot[thick, color=cyan] table [x=ROUND, y=DIFF, col sep=comma] {data/RQ1/loss.csv};
					\addplot[thick, color=magenta] table [x=ROUND, y=HIER, col sep=comma] {data/RQ1/loss.csv};
					\addplot[thick, color=yellow] table [x=ROUND, y=decentral, col sep=comma] {data/RQ1/loss.csv};
				\end{axis}
			\end{tikzpicture}
	}}%
	\subfloat[\centering CPU Usage]{{
			\begin{tikzpicture}
				\begin{axis}[xlabel={Time (s)},ylabel={CPU}, grid=major,grid style={dashed},
					legend pos= north west,legend style={nodes={scale=0.7, transform shape}},
					width=0.333\textwidth,height=4.4cm]
					\addplot[thick, color=orange] table [x=ROUND, y=FEDAVG, col sep=comma] {data/RQ1/CPU.csv};
					\addplot[thick, color=red] table [x=ROUND, y=FEDAVGM, col sep=comma] {data/RQ1/CPU.csv};
					\addplot[thick, color=black] table [x=ROUND, y=SCAFFOLD, col sep=comma] {data/RQ1/CPU.csv};
					\addplot[thick, color=blue] table [x=ROUND, y=MOON, col sep=comma] {data/RQ1/CPU.csv};
					\addplot[thick, color=cyan] table [x=ROUND, y=DIFF, col sep=comma] {data/RQ1/CPU.csv};
					\addplot[thick, color=magenta] table [x=ROUND, y=HIER, col sep=comma] {data/RQ1/CPU.csv};
					\addplot[thick, color=yellow] table [x=ROUND, y=decentral, col sep=comma] {data/RQ1/CPU.csv};
				\end{axis}
			\end{tikzpicture}
	}}%
	\qquad
	\subfloat[\centering Memory Usage]{{
			\begin{tikzpicture}
				\begin{axis}[xlabel={Time (s)},ylabel={Memory (GB)}, grid=major,grid style={dashed},
					legend pos= north west,legend style={nodes={scale=0.7, transform shape}},
					width=0.333\textwidth,height=4.4cm]
					\addplot[thick, color=orange] table [x=ROUND, y=FEDAVG, col sep=comma] {data/RQ1/mem.csv};
					\addplot[thick, color=red] table [x=ROUND, y=FEDAVGM, col sep=comma] {data/RQ1/mem.csv};
					\addplot[thick, color=black] table [x=ROUND, y=SCAFFOLD, col sep=comma] {data/RQ1/mem.csv};
					\addplot[thick, color=blue] table [x=ROUND, y=MOON, col sep=comma] {data/RQ1/mem.csv};
					\addplot[thick, color=cyan] table [x=ROUND, y=DIFF, col sep=comma] {data/RQ1/mem.csv};
					\addplot[thick, color=magenta] table [x=ROUND, y=HIER, col sep=comma] {data/RQ1/mem.csv};
					\addplot[thick, color=yellow] table [x=ROUND, y=decentral, col sep=comma] {data/RQ1/mem.csv};
				\end{axis}
			\end{tikzpicture}
	}}%
	\subfloat[\centering Network Bandwidth]{{
			\begin{tikzpicture}
				\begin{axis}[xlabel={Time (s)},ylabel={Bandwidth (MB)}, grid=major,grid style={dashed},
					legend pos= south east,legend style={at={(2.0,0.0)},nodes={scale=0.7, transform shape}},
					width=0.333\textwidth,height=4.4cm]
					\addplot[thick, color=orange] table [x=ROUND, y=FEDAVG, col sep=comma] {data/RQ1/net.csv};
					\addplot[thick, color=red] table [x=ROUND, y=FEDAVGM, col sep=comma] {data/RQ1/net.csv};
					\addplot[thick, color=black] table [x=ROUND, y=SCAFFOLD, col sep=comma] {data/RQ1/net.csv};
					\addplot[thick, color=blue] table [x=ROUND, y=MOON, col sep=comma] {data/RQ1/net.csv};
					\addplot[thick, color=cyan] table [x=ROUND, y=DIFF, col sep=comma] {data/RQ1/net.csv};
					\addplot[thick, color=magenta] table [x=ROUND, y=HIER, col sep=comma] {data/RQ1/net.csv};
					\addplot[thick, color=yellow] table [x=ROUND, y=decentral, col sep=comma] {data/RQ1/net.csv};
					\legend {FedAvg\cite{fedavg_paper},FedAvgM\cite{fedavgm},SCAFFOLD\cite{scaffold},MOON\cite{moon},Geyer et al. \cite{client_diffpriv},Briggs et al. \cite{hierarchical-cluster},Fedstellar \cite{fedstellar}}
				\end{axis}
			\end{tikzpicture}
	}}%
	\caption{Comparison among state-of-the-art FL techniques}%
	\label{fig:rq1}%
\end{figure}

The detailed evaluation results are shown in Figure \ref{fig:rq1}, showcasing the model metrics as well as resource usage for each of the FL frameworks under consideration. In plot \ref{fig:rq1}a, we observe that the accuracy of MOON \cite{moon} and SCAFFOLD \cite{scaffold} is at the highest, followed by Fedstellar \cite{fedstellar}, FedAvgM \cite{fedavgm}, FedAvg \cite{fedavg_paper}, and \cite{client_diffpriv}, while \cite{hierarchical-cluster} having the lowest overall performance. Similarly, figure \ref{fig:rq1}b depicts the loss curve of each of the FL frameworks, with SCAFFOLD \cite{scaffold} attaining the lowest overall loss between the 10th and 15th rounds. Also, from the plots \ref{fig:rq1}(c,d,e), we can observe that \cite{hierarchical-cluster} requires the most time to complete 30 rounds of FL training, MOON \cite{moon} along with \cite{client_diffpriv,fedavg_paper} requires the least network bandwidth, while Fedstellar with its decentralized peer-to-peer architecture requires the most bandwidth. Based on the information obtained, we can verify that our proposed platform is able to cater to the diverse requirements of these FL frameworks, such as extra state management, additional communication and custom training loops. Additionally, it is able to support the most recent proposal, based on the decentralized FL architecture, i.e., \cite{fedstellar}.

\subsection{\textbf{RQ2}: Evaluation on Different ML Libraries}
We evaluate how the three popular Python ML libraries -- PyTorch, Tensorflow, and Scikit-Learn -- perform on \textsf{FLsim}. All three implementations are tested on the CIFAR-10 dataset distributed using the Dirichlet Distribution algorithm, with $\alpha=0.5$. A total of 10 clients were involved, with batch size and learning rate set to 64 and 0.001. The deep learning model for PyTorch and Tensorflow is comprised of three CNN layers and a fully connected classification head. Since Scikit-Learn does not officially have support for CNN layers, we flatten the CIFAR-10 images and pass them through a Multi-layer Perceptron (MLP) model with four hidden layers. 

\begin{figure}[ht]%
	\centering
	\subfloat[\centering Accuracy]{{
			\begin{tikzpicture}
				\begin{axis}[xlabel={Rounds},ylabel={Accuracy}, grid=major,grid style={dashed},
					legend pos= south east,legend style={nodes={scale=0.7, transform shape}},
					width=0.333\textwidth,height=4.4cm]
					\addplot[thick, color=blue] table [x=round, y=torch, col sep=comma] {data/RQ2/acc.csv};
					\addplot[thick, color=red] table [x=round, y=tf, col sep=comma] {data/RQ2/acc.csv};
					\addplot[thick, color=orange] table [x=round, y=skl, col sep=comma] {data/RQ2/acc.csv};
				\end{axis}
			\end{tikzpicture}
	}}%
	\subfloat[\centering Loss]{{
			\begin{tikzpicture}
				\begin{axis}[xlabel={Rounds},ylabel={Loss}, grid=major,grid style={dashed},
					legend pos= south east,legend style={nodes={scale=0.7, transform shape}},
					width=0.333\textwidth,height=4.4cm]
					\addplot[thick, color=blue] table [x=round, y=torch, col sep=comma] {data/RQ2/loss.csv};
					\addplot[thick, color=red] table [x=round, y=tf, col sep=comma] {data/RQ2/loss.csv};
					\addplot[thick, color=orange] table [x=round, y=skl, col sep=comma] {data/RQ2/loss.csv};
				\end{axis}
			\end{tikzpicture}
	}}%
	\subfloat[\centering CPU Usage]{{
			\begin{tikzpicture}
				\begin{axis}[xlabel={Time (s)},ylabel={CPU}, grid=major,grid style={dashed},
					legend pos= north east,legend style={nodes={scale=0.7, transform shape}},
					width=0.333\textwidth,height=4.4cm]
					\addplot[thick, color=blue] table [x=round, y=torch, col sep=comma] {data/RQ2/cpu.csv};
					\addplot[thick, color=red] table [x=round, y=tf, col sep=comma] {data/RQ2/cpu.csv};
					\addplot[thick, color=orange] table [x=round, y=skl, col sep=comma] {data/RQ2/cpu.csv};
				\end{axis}
			\end{tikzpicture}
	}}%
	\qquad
	\subfloat[\centering Memory Usage]{{
			\begin{tikzpicture}
				\begin{axis}[xlabel={Time (s)},ylabel={Memory (GB)}, grid=major,grid style={dashed},
					legend pos= north east,legend style={nodes={scale=0.7, transform shape}},
					width=0.333\textwidth,height=4.4cm]
					\addplot[thick, color=blue] table [x=round, y=torch, col sep=comma] {data/RQ2/mem.csv};
					\addplot[thick, color=red] table [x=round, y=tf, col sep=comma] {data/RQ2/mem.csv};
					\addplot[thick, color=orange] table [x=round, y=skl, col sep=comma] {data/RQ2/mem.csv};
				\end{axis}
			\end{tikzpicture}
	}}%
	\subfloat[\centering Network Bandwidth]{{
			\begin{tikzpicture}
				\begin{axis}[xlabel={Time (s)},ylabel={Bandwidth (MB)}, grid=major,grid style={dashed},
					legend pos= south east,legend style={at={(1.7,0.0)},nodes={scale=0.7, transform shape}},
					width=0.333\textwidth,height=4.4cm]
					\addplot[thick, color=blue] table [x=round, y=torch, col sep=comma] {data/RQ2/net.csv};
					\addplot[thick, color=red] table [x=round, y=tf, col sep=comma] {data/RQ2/net.csv};
					\addplot[thick, color=orange] table [x=round, y=skl, col sep=comma] {data/RQ2/net.csv};
					\legend {PyTorch,Tensorflow,Scikit-Learn}
				\end{axis}
			\end{tikzpicture}
	}}%
	\caption{Comparison among different ML Libraries}%
	\label{fig:rq2}%
\end{figure}
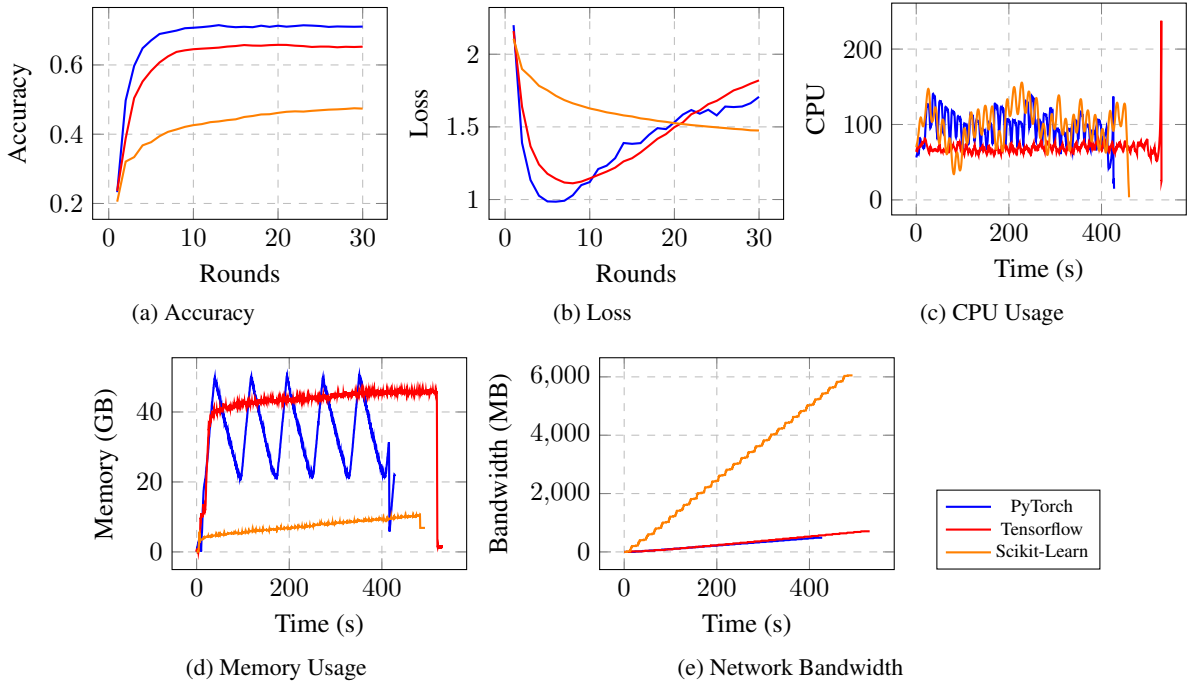

Figure \ref{fig:rq2} details the performance metrics and resource usage for all three ML libraries during the experiments. Based on the results depicted in Plot \ref{fig:rq2}a, the PyTorch implementation achieved the highest accuracy, while the Scikit-Learn-based implementation achieved the lowest due to the different model architecture. Additionally, from Plots \ref{fig:rq2}(c,d,e), we can observe that the Tensorflow implementation required the most time to complete 30 rounds of FL training, and the PyTorch-based implementation required the least. Additionally, the memory usage of the Tensorflow and Scikit-Learn implementation doesn't fluctuate every round and keeps growing gradually, unlike the PyTorch implementation. Also, as depicted in Plot \ref{fig:rq2}e, the Scikit-Learn-based MLP model requires the highest communication bandwidth.

\subsection{\textbf{RQ3}: Evaluation in Multi-Worker Aggregation}

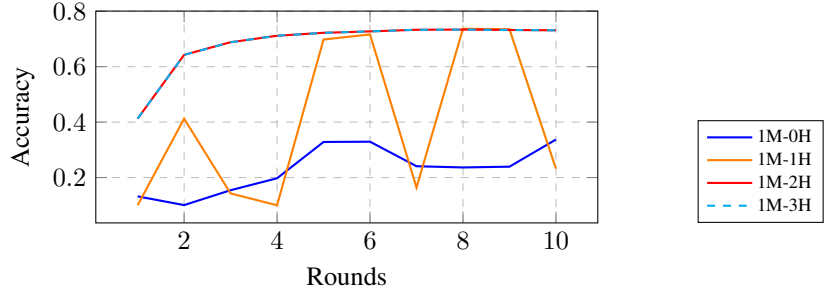
\begin{figure}[ht]%
	\centering
	\subfloat[\centering Accuracy]{{
			\begin{tikzpicture}
				\begin{axis}[xlabel={Rounds},ylabel={Accuracy}, grid=major,grid style={dashed},
					legend pos= south east,legend style={at={(1.45,0.0)},nodes={scale=0.7, transform shape}},
					width=0.5\columnwidth,height=4.4cm]
					\addplot[thick, color=blue] table [x=round, y=1b, col sep=comma] {data/RQ3/acc.csv};
					\addplot[thick, color=orange] table [x=round, y=1b1g, col sep=comma] {data/RQ3/acc.csv};
					\addplot[thick, color=red] table [x=round, y=1b2g, col sep=comma] {data/RQ3/acc.csv};
					\addplot[thick, dashed, color=cyan] table [x=round, y=1b3g, col sep=comma] {data/RQ3/acc.csv};
					\legend {1M-0H,1M-1H,1M-2H,1M-3H}
				\end{axis}
			\end{tikzpicture}
	}}%
	\caption{Malicious Worker Scenario (M = Malicious Worker, H = Honest Worker)}%
	\label{fig:rq3}%
\end{figure}

Here, we demonstrate the support for multi-worker aggregation for FL using \textsf{FLsim}, along with the support for custom consensus algorithms that ensure malicious workers/aggregators cannot affect the learning trajectory through model poisoning attacks. In this regard, we vary the number of workers to demonstrate the effect of malicious workers on poisoning the global model. To this end, we incorporate the worker consensus presented by Chowdhury et al. in \cite{fedrlchain} to orchestrate the multi-worker aggregation scenario. In this experiment, a total of 10 clients and workers ranging from 1 to 4 were involved. The experiment is implemented on the PyTorch library and performed on the CIFAR-10 dataset distributed using the Dirichlet Distribution algorithm, with $\alpha=0.5$. The deep learning model is comprised of three CNN layers and a fully-connected classification head. The batch size was set to 64, and a learning rate of 0.001 were used. We present a scenario of a single malicious worker (1M-0H), one malicious and one honest worker (1M-1H), one malicious and two honest workers (1M-2H), and one malicious and three honest workers (1M-3H). Based on the consensus algorithm in \cite{fedrlchain}, we can observe in Figure \ref{fig:rq3} that when honest clients are $>50\%$ of the total workers, the consensus among workers nullifies the efforts of the malicious worker in poisoning the global model. However, with a 1:1 distribution of malicious and honest workers, a mix of poisoned and healthy global models makes the overall learning process fluctuate in performance.  

\subsection{\textbf{RQ5}: Evaluation under Different Topologies}

\begin{figure}[ht]%
	\centering
	\subfloat[\centering Accuracy]{{
			\begin{tikzpicture}
				\begin{axis}[xlabel={Rounds},ylabel={Accuracy}, grid=major,grid style={dashed},
					legend pos= south east,legend style={nodes={scale=0.7, transform shape}},
					width=0.333\textwidth,height=4.4cm]
					\addplot[thick, color=orange] table [x=round, y=central, col sep=comma] {data/RQ5/acc.csv};
					\addplot[thick, color=cyan] table [x=round, y=hier, col sep=comma] {data/RQ5/acc.csv};
					\addplot[thick, color=magenta] table [x=round, y=decentral, col sep=comma] {data/RQ5/acc.csv};
				\end{axis}
			\end{tikzpicture}
	}}%
	\subfloat[\centering Loss]{{
			\begin{tikzpicture}
				\begin{axis}[xlabel={Rounds},ylabel={Loss}, grid=major,grid style={dashed},
					legend pos= south east,legend style={nodes={scale=0.7, transform shape}},
					width=0.333\textwidth,height=4.4cm]
					\addplot[thick, color=orange] table [x=round, y=central, col sep=comma] {data/RQ5/loss.csv};
					\addplot[thick, color=cyan] table [x=round, y=hier, col sep=comma] {data/RQ5/loss.csv};
					\addplot[thick, color=magenta] table [x=round, y=decentral, col sep=comma] {data/RQ5/loss.csv};
				\end{axis}
			\end{tikzpicture}
	}}%
	\subfloat[\centering CPU]{{
			\begin{tikzpicture}
				\begin{axis}[xlabel={Time (s)},ylabel={CPU}, grid=major,grid style={dashed},
					legend pos= south east,legend style={nodes={scale=0.7, transform shape}},
					width=0.333\textwidth,height=4.4cm]
					\addplot[thick, color=orange] table [x=round, y=central, col sep=comma] {data/RQ5/cpu.csv};
					\addplot[thick, color=cyan] table [x=round, y=hier, col sep=comma] {data/RQ5/cpu.csv};
					\addplot[thick, color=magenta] table [x=round, y=decentral, col sep=comma] {data/RQ5/cpu.csv};
				\end{axis}
			\end{tikzpicture}
	}}%
	\qquad
	\subfloat[\centering Memory Usage]{{
			\begin{tikzpicture}
				\begin{axis}[xlabel={Time (s)},ylabel={Memory (GB)}, grid=major,grid style={dashed},
					legend pos= south east,legend style={nodes={scale=0.7, transform shape}},
					width=0.333\textwidth,height=4.4cm]
					\addplot[thick, color=orange] table [x=round, y=central, col sep=comma] {data/RQ5/mem.csv};
					\addplot[thick, color=cyan] table [x=round, y=hier, col sep=comma] {data/RQ5/mem.csv};
					\addplot[thick, color=magenta] table [x=round, y=decentral, col sep=comma] {data/RQ5/mem.csv};
				\end{axis}
			\end{tikzpicture}
	}}%
	\subfloat[\centering Network Bandwidth]{{
			\begin{tikzpicture}
				\begin{axis}[xlabel={Time (s)},ylabel={Bandwidth (MB)}, grid=major,grid style={dashed},
					legend pos= south east,legend style={at={(2.1,0.0)},nodes={scale=0.7, transform shape}},
					width=0.333\textwidth,height=4.4cm]
					\addplot[thick, color=orange] table [x=round, y=central, col sep=comma] {data/RQ5/net-s.csv};
					\addplot[thick, color=cyan] table [x=round, y=hier, col sep=comma] {data/RQ5/net-s.csv};
					\addplot[thick, color=magenta] table [x=round, y=decentral, col sep=comma] {data/RQ5/net-s.csv};
					\legend {FedAvg \cite{fedavg_paper} (client-server),Briggs et al. \cite{hierarchical-cluster} (hierarchical),Fedstellar \cite{fedstellar} (decentralized)}
				\end{axis}
			\end{tikzpicture}
	}}%
	\caption{Comparison between Client-Server, Hierarchical and Decentralized Topologies}%
	\label{fig:rq5}%
\end{figure}
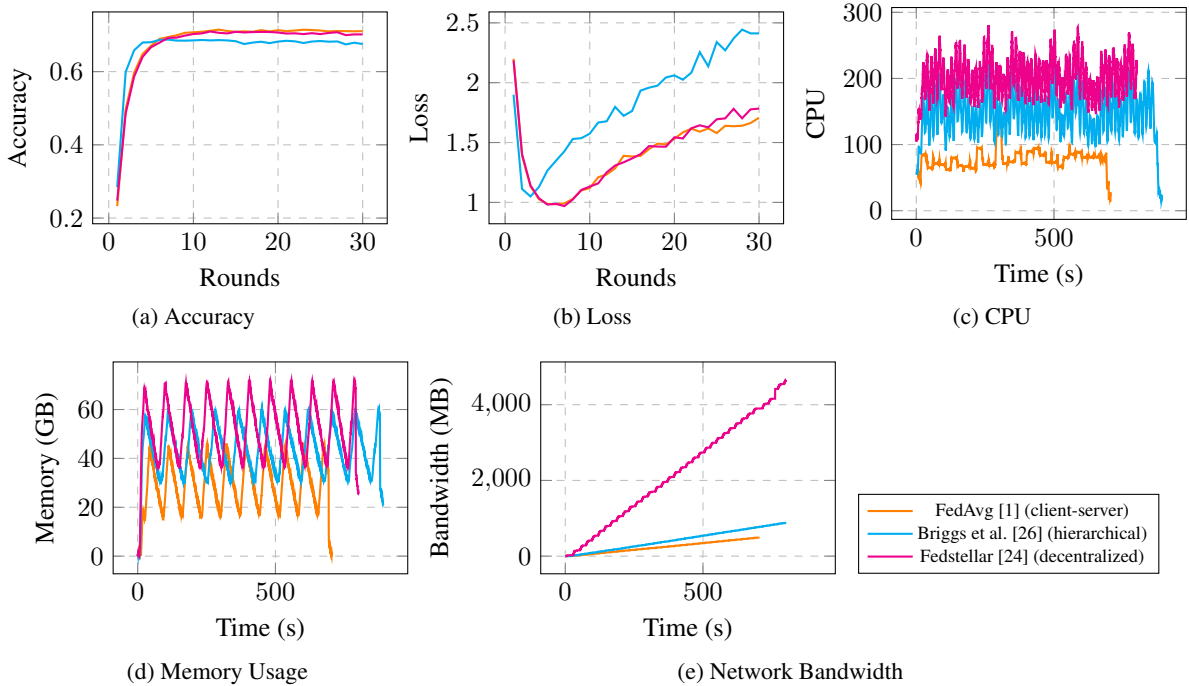

Now, let us manifest the experimental evaluation we performed to assess the support for client-server, hierarchical, and decentralized federated learning on \textsf{FLsim}. The three state-of-the-art proposals we consider for this experiment are as follows: (1) for client-server topology, we consider the simple FedAvg \cite{fedavg_paper} algorithm, (2) for hierarchical topology, we consider \cite{hierarchical-cluster}, and (3) for decentralized FL topology, we consider Fedstellar \cite{fedstellar}. The standard setting is chosen for the experiments, similar to the previous experiments. 
As the information depicted in Figure \ref{fig:rq5}, we can observe that the accuracy of all three topologies is similar. However, the loss for the hierarchical approach is a little higher compared to the client-server and decentralized topology. Additionally, due to the extra computational load of the hierarchical and decentralized topologies, we can observe they have higher CPU and memory usage compared to client-server topology. Also, due to the peer-to-peer communication in decentralized topology, the network bandwidth usage in the case of decentralized topology is the highest.

\subsection{\textbf{RQ6}: Evaluation of Reproducibility}
Since reproducibility is an important factor when it comes to machine learning experiments, in order to gauge the effect of hyperparameters, models and datasets, a core requirement when designing our platform was to enable improved reproducibility capabilities when simulating FL experiments. In this regard, we perform a series of experiments on multiple hardware configurations. All of the experiments were performed with a total of ten clients. The experiment was implemented on the PyTorch library and performed on the CIFAR-10 dataset distributed using the Dirichlet Distribution algorithm, with $\alpha=0.5$. The deep learning model is comprised of three CNN layers and a fully-connected classification head. The batch size was set to 64, and a learning rate of 0.001 was used. The accuracy and loss for the first ten rounds of FL training are recorded for comparison. We performed the first experiment on the CPU of a single x86-based machine running the Intel Xeon Silver 4216 CPU. Next, we run the experiments on the CPU of three x86-based machines (Intel Xeon Silver 4216, Intel Xeon E5-4650 v4, and Intel Core i5-10500) in a 5-3-2 distributed setup. The third experiment is performed on an NVIDIA A100 GPU with an x86-based processor (Intel Xeon Silver 4216 CPU). Finally, we run our experiments on the CPU of an ARM platform (aarch64), i.e., a Raspberry Pi 4.

\begin{table*}[ht]
	\centering
	\footnotesize
	\caption{Reproducibility Results for Accuracy}
	\label{tab:reproducibility-acc}
	\begin{tabular}{cccccccccccc}
		\hline
		& & \multicolumn{10}{c}{Accuracy at FL Round} \\ \cline{3-12} 
		\multirow{-2}{*}{{\wrap{Type}}}  & \multirow{-2}{*}{Trial} & 1 & 2 & 3 & 4 & 5 & 6 & 7 & 8 & 9 & 10 \\ \hline
		x86 Single CPU &  & 0.4433 & 0.5575 & 0.6100 & 0.6446 & 0.6716 & 0.6887 & 0.6989 & 0.7023 & 0.7168 & 0.7189 \\
		x86 Dist CPU & & 0.4409 & 0.5537 & 0.6096 & 0.6456 & 0.6709 & 0.6854 & 0.6966 & 0.7033 & 0.7128 & 0.7162 \\
		x86 Single GPU & & 0.4109 & 0.5532 & 0.6121 & 0.6482 & 0.6731 & 0.6885 & 0.6998 & 0.7068 & 0.7186 & 0.7223 \\
		aarch64 Single CPU & \multirow{-4}{*}{1} & 0.4365 & 0.5545 & 0.611 & 0.6468 & 0.6737 & 0.6883 & 0.7014 & 0.7042 & 0.7181 & 0.7202 \\ \hline
		x86 Single CPU &  & 0.4433 & 0.5575 & 0.6100 & 0.6446 & 0.6716 & 0.6887 & 0.6989 & 0.7023 & 0.7168 & 0.7189 \\
		x86 Dist CPU & & 0.4409 & 0.5537 & 0.6096 & 0.6456 & 0.6709 & 0.6854 & 0.6966 & 0.7033 & 0.7128 & 0.7162 \\
		x86 Single GPU & & 0.4109 & 0.5532 & 0.6121 & 0.6482 & 0.6731 & 0.6885 & 0.6998 & 0.7068 & 0.7186 & 0.7223 \\
		aarch64 Single CPU & \multirow{-4}{*}{2} & 0.4365 & 0.5545 & 0.611 & 0.6468 & 0.6737 & 0.6883 & 0.7014 & 0.7042 & 0.7181 & 0.7202 \\ \hline
		x86 Single CPU &  & 0.4433 & 0.5575 & 0.6100 & 0.6446 & 0.6716 & 0.6887 & 0.6989 & 0.7023 & 0.7168 & 0.7189 \\
		x86 Dist CPU & & 0.4409 & 0.5537 & 0.6096 & 0.6456 & 0.6709 & 0.6854 & 0.6966 & 0.7033 & 0.7128 & 0.7162 \\
		x86 Single GPU & & 0.4109 & 0.5532 & 0.6121 & 0.6482 & 0.6731 & 0.6885 & 0.6998 & 0.7068 & 0.7186 & 0.7223 \\
		aarch64 Single CPU & \multirow{-4}{*}{3} & 0.4365 & 0.5545 & 0.611 & 0.6468 & 0.6737 & 0.6883 & 0.7014 & 0.7042 & 0.7181 & 0.7202 \\ \hline
	\end{tabular}
\end{table*}

\begin{table*}[ht]
	\centering
	\footnotesize
	\caption{Reproducibility Results for Loss}
	\label{tab:reproducibility-loss}
	\begin{tabular}{cccccccccccc}
		\hline
		& & \multicolumn{10}{c}{Loss at FL Round} \\ \cline{3-12} 
		\multirow{-2}{*}{{\wrap{Type}}}  & \multirow{-2}{*}{Trial} & 1 & 2 & 3 & 4 & 5 & 6 & 7 & 8 & 9 & 10 \\ \hline
		x86 Single CPU &  & 1.6083 & 1.2396 & 1.0930 & 1.0014 & 0.9349 & 0.8909 & 0.8708 & 0.8825 & 0.8419 & 0.8674 \\
		x86 Dist CPU & & 1.6165 & 1.2422 & 1.0929 & 1.0008 & 0.9358 & 0.8948 & 0.8669 & 0.8707 & 0.8450 & 0.8668 \\
		x86 Single GPU & & 1.6245 & 1.2509 & 1.0987 & 1.0015 & 0.9392 & 0.8899 & 0.8714 & 0.8794 & 0.8435 & 0.8648 \\
		aarch64 Single CPU & \multirow{-4}{*}{1} & 1.6138 & 1.2431 & 1.0930 & 1.0002 & 0.9364 & 0.8887 & 0.8698 & 0.8773 & 0.8428 & 0.8687 \\ \hline
		x86 Single CPU &  & 1.6083 & 1.2396 & 1.0930 & 1.0014 & 0.9349 & 0.8909 & 0.8708 & 0.8825 & 0.8419 & 0.8674 \\
		x86 Dist CPU & & 1.6165 & 1.2422 & 1.0929 & 1.0008 & 0.9358 & 0.8948 & 0.8669 & 0.8707 & 0.8450 & 0.8668 \\
		x86 Single GPU & & 1.6245 & 1.2509 & 1.0987 & 1.0015 & 0.9392 & 0.8899 & 0.8714 & 0.8794 & 0.8435 & 0.8648 \\
		aarch64 Single CPU & \multirow{-4}{*}{2} & 1.6138 & 1.2431 & 1.0930 & 1.0002 & 0.9364 & 0.8887 & 0.8698 & 0.8773 & 0.8428 & 0.8687 \\ \hline
		x86 Single CPU &  & 1.6083 & 1.2396 & 1.0930 & 1.0014 & 0.9349 & 0.8909 & 0.8708 & 0.8825 & 0.8419 & 0.8674 \\
		x86 Dist CPU & & 1.6165 & 1.2422 & 1.0929 & 1.0008 & 0.9358 & 0.8948 & 0.8669 & 0.8707 & 0.8450 & 0.8668 \\
		x86 Single GPU & & 1.6245 & 1.2509 & 1.0987 & 1.0015 & 0.9392 & 0.8899 & 0.8714 & 0.8794 & 0.8435 & 0.8648 \\
		aarch64 Single CPU & \multirow{-4}{*}{3} & 1.6138 & 1.2431 & 1.0930 & 1.0002 & 0.9364 & 0.8887 & 0.8698 & 0.8773 & 0.8428 & 0.8687 \\ \hline
	\end{tabular}
\end{table*}

The results obtained from the four experiments executed on three separate trials are reported in Tables \ref{tab:reproducibility-acc} and \ref{tab:reproducibility-loss}. As we can observe, the accuracy and loss for the three separate trials on the same hardware configurations yield the exact same numbers. In contrast, the results from the four different hardware configurations yield a slightly different result, with accuracies varying up to 0.6\% at the tenth round. This slight variation in performance on different hardware configurations is due to the different hardware-level implementations and variations in the floating-point arithmetic involved in machine learning. A detailed discussion on why such variations take place is discussed in Section \ref{sec:rel-work-discussion}.

\subsection{\textbf{RQ7}: Large-scale Experiments}

The scalability of an FL simulation framework is essential since systems might scale to hundreds to thousands of clients in reality. In this regard, we perform four experiments using the MNIST dataset on 100, 250, 500, and 1000 clients, implemented on the Scikit-Learn library.

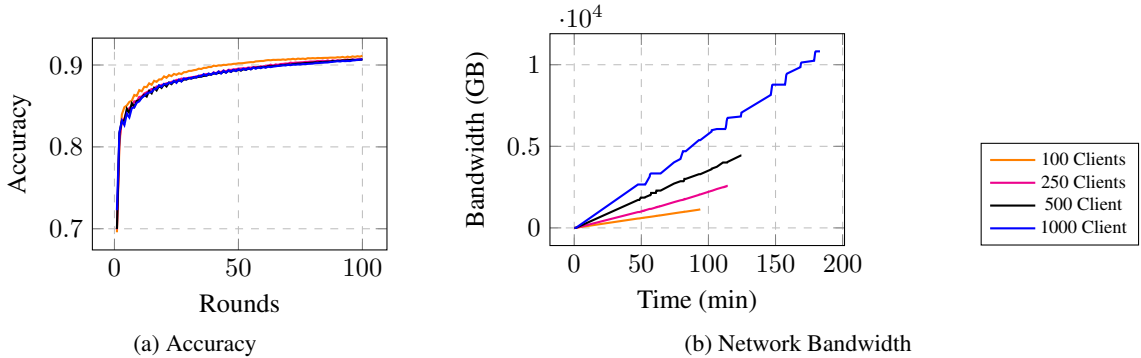
\begin{figure}[ht]%
	\centering
	\subfloat[\centering Accuracy]{{
			\begin{tikzpicture}
				\begin{axis}[xlabel={Rounds},ylabel={Accuracy}, grid=major,grid style={dashed},
					legend pos= south east,legend style={nodes={scale=0.7, transform shape}},
					width=0.333\textwidth,height=4.4cm]
					\addplot[thick, color=orange] table [x=ROUND, y=100C, col sep=comma] {data/RQ7/acc.csv};
					\addplot[thick, color=magenta] table [x=ROUND, y=250C, col sep=comma] {data/RQ7/acc.csv};
					\addplot[thick, color=black] table [x=ROUND, y=500C, col sep=comma] {data/RQ7/acc.csv};
					\addplot[thick, color=blue] table [x=ROUND, y=1000C, col sep=comma] {data/RQ7/acc.csv};
				\end{axis}
			\end{tikzpicture}
	}}%
	\qquad
	\subfloat[\centering Network Bandwidth]{{
			\begin{tikzpicture}
				\begin{axis}[xlabel={Time (min)},ylabel={Bandwidth (GB)}, grid=major,grid style={dashed},
					legend pos= south east,legend style={at={(2.0,0.0)},nodes={scale=0.7, transform shape}},
					width=0.333\textwidth,height=4.4cm]
					\addplot[thick, color=orange] table [x=TIME, y=100C, col sep=comma] {data/RQ7/net-s.csv};
					\addplot[thick, color=magenta] table [x=TIME, y=250C, col sep=comma] {data/RQ7/net-s.csv};
					\addplot[thick, color=black] table [x=TIME, y=500C, col sep=comma] {data/RQ7/net-s.csv};
					\addplot[thick, color=blue] table [x=TIME, y=1000C, col sep=comma] {data/RQ7/net-s.csv};
					\legend {100 Clients,250 Clients,500 Client,1000 Client}
				\end{axis}
			\end{tikzpicture}
	}}%
	\caption{Large-scale Experimental Results on MNIST}%
	\label{fig:rq7}%
\end{figure}

The experiments performed on the MNIST dataset comprised a logistic regression model, with the dataset being distributed uniformly among 100, 250, 500, and 1000 clients. The results of the experiments are depicted in Figure \ref{fig:rq7}. As evident from the experiments, we can observe that the performance of the global model (accuracy) is the same overall for all four client settings. However, it is evident from the plot of Network Bandwidth, i.e., Figure \ref{fig:rq7}b, the network bandwidth usage increases with the increase in the number of clients. Additionally, the overall time required to complete the set number of rounds increases from 100 to 1000 clients.

\section{Related Work \& Discussion}\label{sec:rel-work-discussion}

In recent years, a number of initiatives have emerged to advance federated learning through the development of specialized libraries tailored to diverse testing and deployment scenarios. Notable among these are simulation-oriented platforms like TensorFlow Federated (TFF) \cite{tff} and PySyft \cite{pysyft}, which primarily focus on facilitating federated learning simulations. Conversely, platforms like FATE \cite{fate} and FedBioMed \cite{fedbiomed} are directed towards real-world deployment scenarios. Flower \cite{flower} stands out for its user-friendly integration of exploratory research in federated learning, providing support for modular FL algorithms and scalability across heterogeneous devices. Following Flower's lead, subsequent platforms like FedLab \cite{fedlab} and FedML \cite{fedml} have further contributed to the FL ecosystem. FedLab offers a flexible framework for FL simulations, empowering users with essential functions for model optimization, communication, and data partitioning. In a significant stride towards decentralized FL, Beltrán et al. introduced Fedstellar \cite{fedstellar} in 2023. This platform not only facilitates decentralized FL but also enhances support for existing FL frameworks while providing users with an intuitive frontend for experiment management and progress tracking.

\begin{sidewaystable}
	\centering
	\caption{Comparative analysis among federated learning simulation and testing frameworks}
	\label{tab:comparative-analysis}
	\begin{tabular}{lp{0.7in}p{1.0in}p{0.8in}p{0.7in}p{0.8in}p{0.6in}p{0.8in}p{0.8in}}
		\hline
		Framework & FL & Topology & Libraries & Data & Deployment & Metrics & ML & Blockchain \\
		& Architecture & & & Distribution & & Logging & Framework & Framework \\
		& & & & & & & Agnostic & Agnostic \\ \hline
		FATE \cite{fate} & CFL & Client-server & Tensorflow, PyTorch & \xmark & Production, Simulation & \xmark & \cmark & \xmark \\ \hline
		TFF \cite{tff} & CFL & Client-server & Tensorflow & \xmark & Production, Simulation & \cmark & \xmark & \xmark \\ \hline
		PySyft \cite{pysyft} & CFL & Client-server & PyTorch & \xmark & Production & \cmark & \xmark & \xmark \\  \hline
		FedML \cite{fedml} & CFL, HFL, DFL & Fully-connected & PyTorch & \xmark & Simulation & \cmark & \xmark & \xmark \\  \hline
		Flower \cite{flower} & CFL & Client-server & Tensorflow, PyTorch & \xmark & Production, Simulation & \cmark & \cmark & \xmark \\  \hline
		Fedstellar \cite{fedstellar} & CFL, SDFL, DFL & Fully-connected & PyTorch & \cmark & Production & \cmark & \xmark & \xmark \\  \hline
		FedLab \cite{fedlab} & CFL & Client-server & PyTorch & \cmark & Production, Simulation & \cmark & \xmark & \xmark \\  \hline
		FedBioMed \cite{fedbiomed} & CFL & Client-server & PyTorch, Scikit-Learn & \xmark & Production & \cmark & \cmark & \xmark \\  \hline
		\textsf{FLsim} (Ours) & CFL, HFL, SDFL, DFL & Fully-connected & PyTorch, Tensorflow, Scikit-Learn & \cmark & Simulation & \cmark & \cmark & \cmark \\  \hline
	\end{tabular}
\end{sidewaystable}

A comparative analysis of the state-of-the-art  frameworks w.r.t. \textsf{FLsim} is provided in Table \ref{tab:comparative-analysis}. Observe that, unlike the existing simulation and deployment frameworks, \textsf{FLsim} is flexible enough to support any FL topology, is completely ML library agnostic, and has support for blockchain-based features, such as model parameter verification and global model provenance. Let us now highlight the way we achieve these key features of \textsf{FLsim} that distinguish it from the existing frameworks.

The primary objective of \textsf{FLsim} is to make the simulation framework completely ML library agnostic as well as to enhance its ability to simulate a wide range of FL paradigms. Moreover, the design of \textsf{FLsim} ensures its workflow is simple enough for users so as to adopt any ML library for their use cases. To this aim, we adopt an object-oriented (OO) approach, where everything, the dataset, model, training loop, testing loop, and aggregation function, is baked into this single class definition (i.e., \textsf{FLsim} Strategy). This new OO-based approach enables the complete framework to become more customizable and modular by the likes of the object-oriented programming paradigm and future-proofed the framework's design in terms of expandability for any FL requirement, be it aggregation algorithm-wise, ML model-wise, or communication-wise.

The next challenge that \textsf{FLsim} effectively attempted to address is the reproducibility of FL experiments. As reported in the survey conducted by Baker \cite{repro-survey} in 2016, it is observed that more than 50\% of researchers fail to reproduce the results of their own ML experiments. Distributed settings add more complexities towards reproducibility. In this context, when reproducing results of the same experiment with all the seeds and environment variables set, there comes the enormous challenge of having different CPU architectures and designs. To effectively address these challenges, we use a node seed synchronization technique that enables all of the \textsf{FLsim} nodes to initialize on a set of seed values for controlling the randomness and enabling deterministic execution. Additionally, we define methods to handle and configure the specific ML libraries to enable deterministic execution on their side, such as configuring \texttt{torch.use\_deterministic\_algorithms(True)} in the case of PyTorch. To enable deterministic and reproducible experiments, users need to set the environment variable \texttt{DETERMINISTIC} to \texttt{true} and optionally set the random seed by setting an integer value to the environment variable \texttt{RANDOM\_SEED}.

Further, when scaling the system to a large number of nodes, we noticed that the Logic Controller was under a Distributed Denial of Service (DDoS) attack from the nodes, due to a high number of polling requests from the nodes to the Logic Controller. This was easily circumvented by reducing the polling rate of the nodes. Additionally, we introduced some load balancing for the Logic Controller. This was done by first upgrading the Flask server to use \texttt{waitress}, which uses CPython under the hood. Next, we introduced an NGINX-based load balancing with $n$ instances of Logic Controller working behind the load balancer proxy.

\section{Conclusion \& Future Scope}\label{sec:conclusion}
This paper introduces \textsf{FLsim}, a federated learning (FL) simulation framework that advances upon existing platforms by catering to diverse FL framework requirements. Unlike existing platforms, \textsf{FLsim} enables learning over heterogeneous data distributions among clients while also achieving complete agnosticism from machine learning (ML) libraries to accommodate the preferences of users across different ML frameworks. Moreover, \textsf{FLsim} supports a wide range of network topologies, from client-server to decentralized setups, and ensures controlled reproducibility of experimental outcomes. Scalability is a key feature, allowing \textsf{FLsim} to accommodate a large number of nodes, making it suitable for both small-scale experiments and large-scale FL deployments. Notably, \textsf{FLsim} empowers users with modular and highly customizable configuration options, facilitating the creation of tailored dataset distributions and FL algorithmic strategies. Through extensive experimental evaluation, we validate the practicality and effectiveness of \textsf{FLsim} under various conditions and scenarios. 

As the framework continues to evolve, we are in the process of enhancing the efficiency and memory management of the framework, along with optimizing job scheduling mechanisms to expedite the execution of FL experiments. Further, we are working on expanding our repository of demo experiments and templates for existing FL proposals and diversifying the collection. 

\section*{Acknowledgement}
This research is supported by the Research Grant (File Number: B-13011/1/2022-Training/3134963) from MeitY, NIC, Government of India. The authors further express their gratitude to Sujit Chowdhury, Hemant Chaurasia, Mamta Kanwar, and Akash Sinha for their valuable contributions and feedback to enrich the development and refinement of the work. 

\bibliography{ref.bib}
\bibliographystyle{unsrt}

\end{document}